\RequirePackage{fixltx2e}
\documentclass[a4paper]{isp}
\usepackage[american]{babel}
\usepackage{graphicx}
\usepackage{amsmath,amsthm,mathtools}
\usepackage{url}
\usepackage[capitalise,nameinlink]{cleveref}
\usepackage{xcolor}
\usepackage[utf8]{inputenc}
\usepackage{enumitem}
\def\aap{A\&A\,  }

\def\aaps{A\&AS  }

\def\mnras{MNRAS\,  }

\def\2F1{~_2F_1}
\begin{document}
\pdfgentounicode=1
\title
{
New probability distributions in astrophysics:
IV. The relativistic Maxwell-Boltzmann distribution
}
\author{Lorenzo  Zaninetti}
\institute{
Physics Department,
 via P.Giuria 1, I-10125 Turin,Italy \\
 \email{zaninetti@ph.unito.it}
}

\maketitle

\begin {abstract}
Two relativistic  distributions which generalizes 
the Maxwell Boltzman (MB) distribution are analyzed:
the relativistic MB and the 
Maxwell-J{\"u}ttner (MJ) distribution.
For the two distributions we derived
in terms  of special functions  
the constant of normalization,
the average value,
the second moment about the origin,
the variance,
the mode,
the asymptotic behavior,
approximate expressions for the average value as function of the
temperature and the connected 
inverted expressions for the temperature as function of 
the average value.
Two astrophysical applications to the synchrotron emission 
in presence of the magnetic field and the relativistic electrons
are presented.
\end{abstract}

\section{Introduction}

The equivalent in special relativity (SR) 
of the {\em Maxwell-Boltzmann} (MB) distribution, see
\cite{Maxwell1860,Boltzmann1872},
is the so called {\em Maxwell-J{\"u}ttner} distribution (MJ),
see \cite{Juttner1911,Synge1957}.
The MJ distribution has been recently revisited,
we select some  approaches among others: 
a model for the
anisotropic MJ distribution \cite{Livadiotis2016},
an astrophysical application of the MJ distribution
to the energy distribution in radio jets \cite{Tsouros2017},
a new family of MJ distributions  characterized 
by the parameter $\eta$ \cite{Aragon2018}
and
an  application to counter-streaming beams of charged particles
\cite{Sadegzadeh2018}.
The above approaches does not cover the determination 
of the statistical quantities of the MJ distribution.
In this paper 
the statistical  parameters   of the 
relativistic MB distribution are derived 
in Section \ref{secmb} 
and those 
of the MJ distribution are derived 
in Section \ref{secmj}.
Section \ref{secastro}  derives  the  
spectral synchrotron emissivity  in the framework
of the two relativistic distributions here analyzed.
 
\section{The relativistic MB distribution}

\label{secmb}
The usual MB distribution, $f(v;m,k,T_{MB})$,
for an ideal  gas is
\begin{equation}
f(v;m,k,T_{MB}) =
\frac
{
\sqrt {2}{v}^{2}{{\rm e}^{-\frac{1}{2}\,{\frac {{v}^{2}m}{kT_{MB}}}}}
}
{
\sqrt {\pi} \left( {\frac {kT_{MB}}{m}} \right) ^{\frac{3}{2}}
}
\quad ,
\label{mbdistribution}
\end{equation}
where $m$ is  the mass of the gas molecules,
$k$ is the Boltzmann constant and 
$T_{MB}$ is the usual thermodynamic temperature.
In SR, the total energy of a particle  is
\begin{equation}
E = m \gamma c^2
\quad ,
\end{equation}
where $m$ is the rest mass,
      $c$ is the light velocity,
      $\gamma$ is the Lorentz factor $\frac{1}{\sqrt{1-\beta^2}}$,
      $\beta=\frac{v}{c}$ and $v$ is the velocity.
The relativistic kinetic energy, $E_k$, is
\begin{equation}
E_k = m c^2 (\gamma-1)
\quad ,
\end{equation}
where the rest energy has been subtracted
from the total energy, see formula (23.1) in \cite{Freund2008}.
A relativistic  MB distribution can be obtained from
equation (\ref{mbdistribution}) replacing 
the classical  kinetic energy $\frac{1}{2} m v^2$ 
with the relativistic kinetic energy 
\begin{equation}
f_r(v;T) =
\frac
{
{v}^{2}{{\rm e}^{{\frac {1}{T} \left( 1-{\frac {1}{\sqrt {1-\frac{{v}^{2}}
{c^2}
}}
} \right) }}}
}
{
\int_{0}^{c}\!{w}^{2}{{\rm e}^{{\frac {1}{T} \left( 1-{\frac {1}{
\sqrt {1-\frac{{w}^{2}}
{c^2}
}}} \right) }}}\,{\rm d}w
}
\quad ,
\label{mbrelativistic}
\end{equation}
where the relativistic temperature, $T$, is expressed in $m\,c^2/k$ 
units;
up to now the treatment is the same of \cite{Claycomb2018}
at pag.~665.
The above relativistic PDF 
 \setlist{nolistsep}
\begin{itemize}
\item 
has the velocity of the light as maximum velocity,
\item 
becomes the  usual MB distribution in the limit of low velocities,
\item
is not invariant for relativistic transformations.
\end{itemize}

\subsection{Variable Lorentz factor}

\begin{figure*}
\begin{center}
\includegraphics[width=5cm]{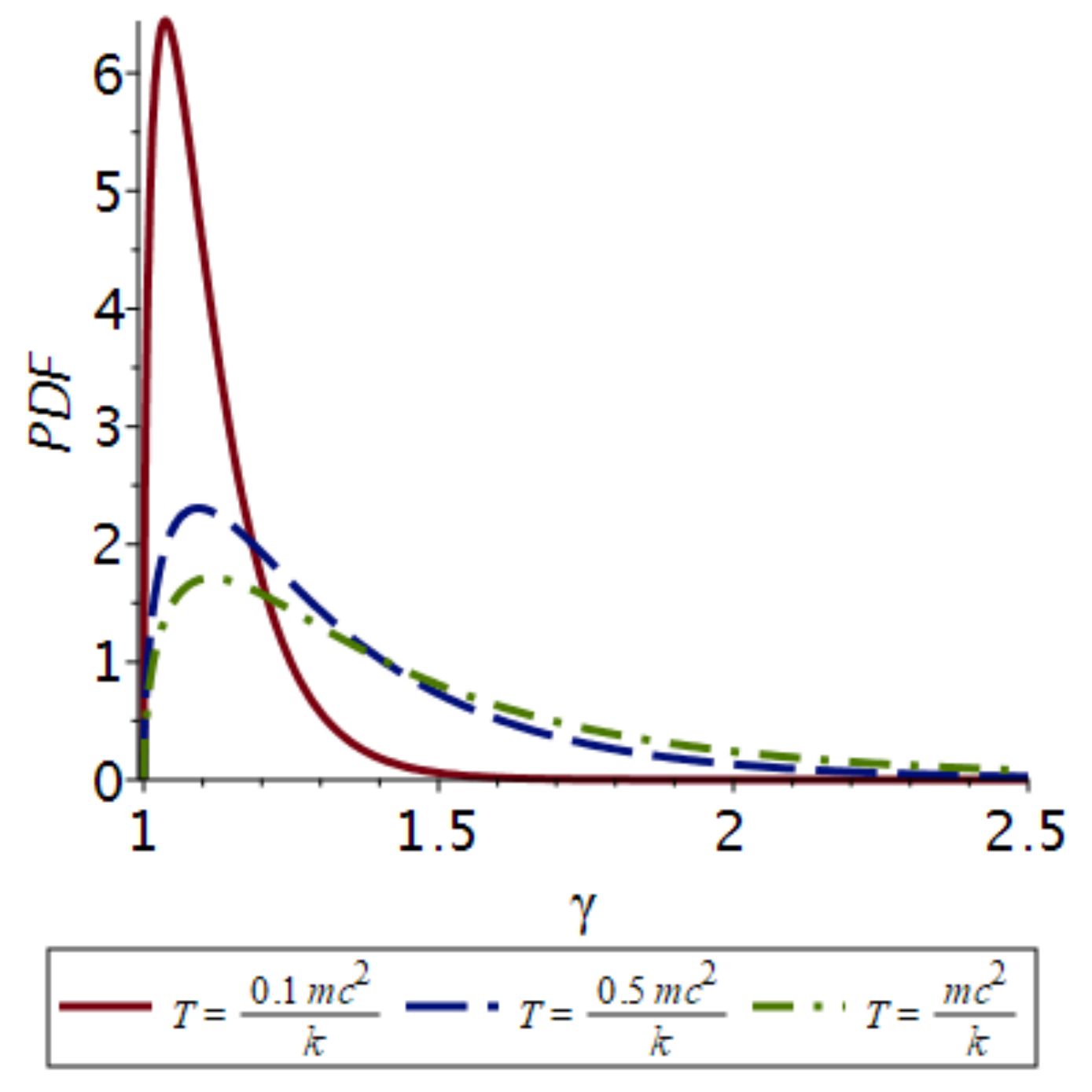}
\end {center}
\caption
{
The PDF of the relativistic MB as function of
$\gamma$ for different values of T.
}
\label{pdfgamma}
    \end{figure*}

We now change the variable of integration
\begin{equation}
v={\frac {\sqrt {{\gamma}^{2}-1}}{\gamma}}
\quad .
\end{equation}
The differential of the velocity, $dv$,
\begin{equation}
dv = {\frac {1}{\sqrt {{\gamma}^{2}-1}{\gamma}^{2}}} d\gamma
\quad ,
\end{equation} 
and therefore 
the relativistic MB distribution in the variable $\gamma$ is
\begin{equation}
f_r(\gamma;T)=
\frac
{
32\,\sqrt {{\gamma}^{2}-1}{{\rm e}^{{\frac {1-\gamma}{T}}}}{T}^{3}
}
{
{\gamma}^{4}{{\rm e}^{{T}^{-1}}}
G^{3, 0}_{1, 3}\left(\frac{1}{4T^2}\, \Big\vert\,^{1}_{-1/2, -1, -3/2}\right)
}
\quad ,
\label{pdfgammarel}
\end{equation}
where  $G$ is the Mejier  $G$-function
\cite{Meijer1936,Meijer1941,NIST2010};
Figure \ref{pdfgamma}  reports the above PDF
for three different temperatures.

The average  value or mean, $\mu$,  is
\begin{equation}
\mu(T) =
\frac
{
2\,T
G^{3, 0}_{1, 3}\left(\frac{1}{4T^2}\, \Big\vert\,^{1}_{-1/2, -1/2, -1}\right)
}
{
G^{3, 0}_{1, 3}\left(\frac{1}{4T^2}\, \Big\vert\,^{1}_{-1/2, -1, -3/2}\right)
}
\quad ,
\end{equation}
the second  moment about the origin is
\begin{equation}
\mu(T)^2 =
\frac
{
4\,{T}^{2}
G^{3, 0}_{1, 3}\left(1/4\,{T}^{-2}\, \Big\vert\,^{1}_{0, -1/2, -1/2}\right)
}
{
G^{3, 0}_{1, 3}\left(1/4\,{T}^{-2}\, \Big\vert\,^{1}_{-1/2, -1, -3/2}\right)
}
\quad ,
\end{equation}
the variance, $\sigma^2$   is  
\begin{eqnarray}
\sigma^2(T)=
\nonumber \\
\frac
{
4\,{T}^{2} \left( 
G^{3, 0}_{1, 3}\left(\frac{1}{4T^2}\, \Big\vert\,^{1}_{-1/2, -1, -3/2}\right)
G^{3, 0}_{1, 3}\left(\frac{1}{4T^2}\, \Big\vert\,^{1}_{0, -1/2, -1/2}\right)
- \left( 
G^{3, 0}_{1, 3}\left(\frac{1}{4T^2}\, \Big\vert\,^{1}_{-1/2, -1/2, -1}\right)
 \right) ^{2} \right) 
}
{
\left( 
G^{3, 0}_{1, 3}\left(\frac{1}{4T^2}\, \Big\vert\,^{1}_{-1/2, -1, -3/2}\right)
 \right) ^{2}
}
\quad .
\end{eqnarray}
The mode is the real solution of the following cubic  
equation in $\gamma$   
\begin{equation}
{{\it \gamma}}^{3}+3\,T{{\it \gamma}}^{2}-{\it \gamma}-4\,T=0
\quad ,
\end{equation}
which has the real solution
\begin{eqnarray}
mode=\frac{1}{6}\,\sqrt [3]{324\,T-216\,{T}^{3}+12\,\sqrt {-1296\,{T}^{4}+621\,{T}^
{2}-12}}
\nonumber \\
-6\,{\frac {-1/3-{T}^{2}}{\sqrt [3]{324\,T-216\,{T}^{3}+12\,
\sqrt {-1296\,{T}^{4}+621\,{T}^{2}-12}}}}-T
\quad .
\end{eqnarray}

\begin{figure*}
\begin{center}
\includegraphics[width=5cm]{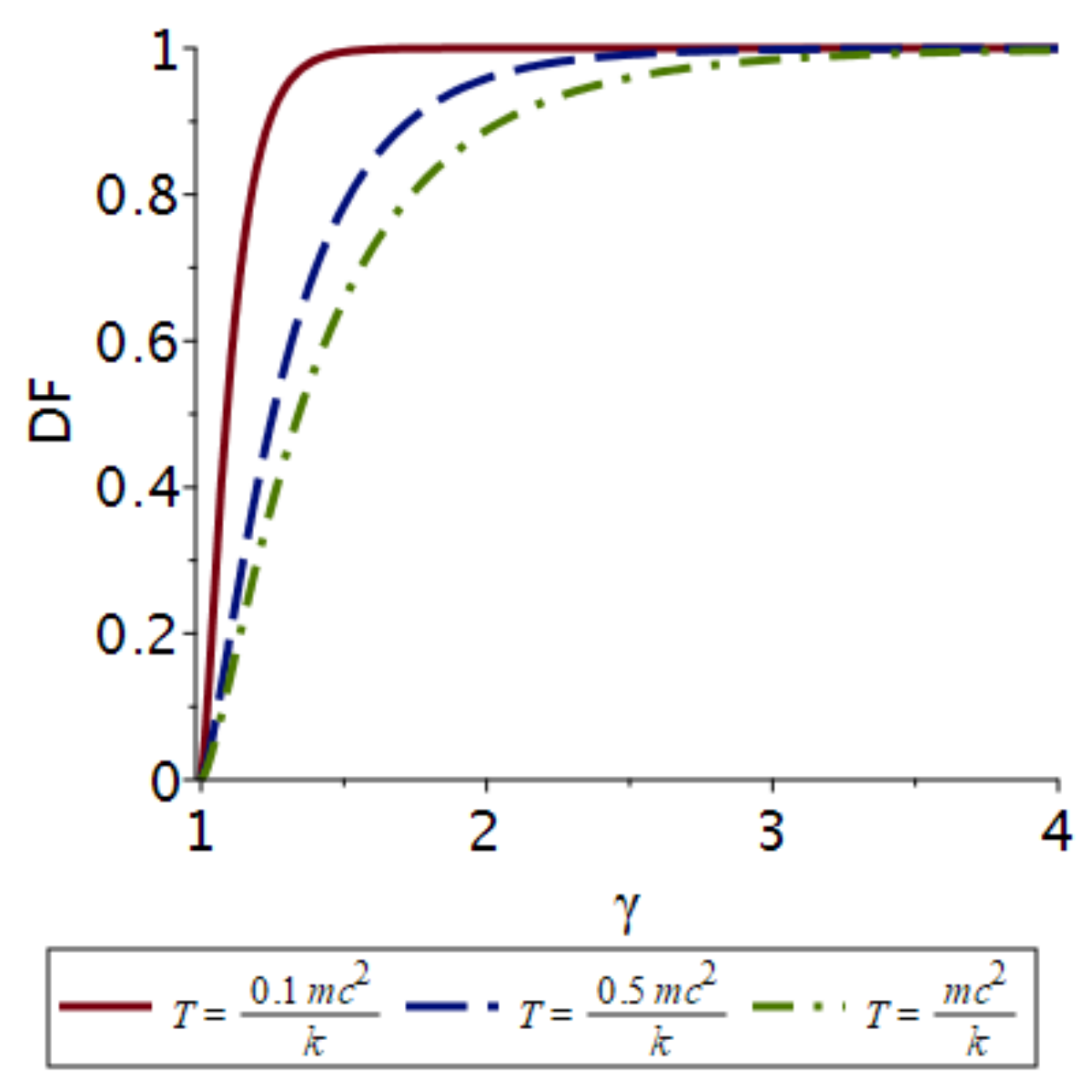}
\end {center}
\caption
{
The numerical DF of the relativistic MB as function of
$\gamma$ for different values of T.
}
\label{dfgamma}
    \end{figure*}

At the moment of writing 
a closed form for the distribution function (DF)
which is 
\begin{equation}
F_r(\gamma;T) = \int_1^y f_r(\gamma;T) d\gamma
\label{dfmaxwellrel}  
\quad ,
\end{equation}  
does not exists  and we therefore present  
a numerical integration, see Figure \ref{dfgamma}.

\begin{figure*}
\begin{center}
\includegraphics[width=5cm,angle=-90]{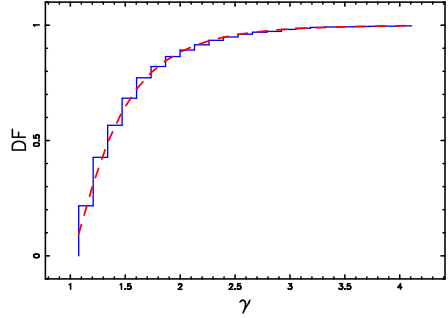}
\end {center}
\caption
{
The DF  for 3000 random points
generated according to the relativistic  MB (blue steps diagram) 
and the theoretical DF (red dashed line)   
see \ref{df_maxwellrel}.
}
\label{df_maxwellrel}
    \end{figure*}

The asymptotic behavior of the PDF, $f_a$, is
\begin{eqnarray}
f_a(\gamma;T)= 
\frac{1}
{
G^{3, 0}_{1, 3}\left(1/4\,{T}^{-2}\,
\Big\vert\,^{1}_{-1/2, -1, -3/2}\right)
\,2048\,{{\it \gamma}}^{21}
}
{T}^{3} \Bigg( 65536\,{{\it \gamma}}^{18}-32768\,{{\it \gamma}}^{16}-
8192\,{{\it \gamma}}^{14}
\nonumber \\
-4096\,{{\it \gamma}}^{12}-2560\,{{\it \gamma
}}^{10}-1792\,{{\it \gamma}}^{8}-1344\,{{\it \gamma}}^{6}-1056\,{{\it 
\gamma}}^{4}-858\,{{\it \gamma}}^{2}-715 \Bigg ) {{\rm e}^{-{\frac {{
\it \gamma}}{T}}}}
\quad .
\end{eqnarray}
The integration of the above approximate PDF 
gives an approximate DF   which 
has a maximum percentage error of $7\%$ in the
interval $1.1 < \gamma <4$ when  $T=1$.
The  random   numbers   belonging
to the  relativistic MB 
can  be  generated
through a numerical computation
of the inverse function following
the algorithm outlined in
Sec. 4.9.1 of  \cite{Brandt1998}.
The above  PDF has only one parameter which can be derived  
approximating the average value  with  a Pade approximant $[2,2]$
\begin{equation}
\mu(T) \approx 
{\frac {- 0.061723842+ 1.542917977\,T+ 0.3269078746\, \left( T- 1
 \right) ^{2}}{ 0.1069596119+ 0.8930403881\,T+ 0.1511024609\, \left( T
- 1 \right) ^{2}}}
\quad .
\end{equation}
The above approximation in the 
interval $0.1 \leq T < 10$ has a  percent error less than 1$\%$.
The inverse function allows to derive $T$ as
\begin{equation}
T=-\frac{1}{2}\,{\frac { 5.908 \times10^9\,{\it \bar{x}}- 8.89\times10^9  +\sqrt 
{{ 1.931\times 10^{19}}\,{{\it \bar{x}}}^{2}-{ 5.528 \times 10
^{19}}\,{\it \bar{x}}+{ 4.437 \times 10^{19}}}}{ 1.511 \times10^9 \,{\it 
\bar{x}}- 3.269\times10^9}}
\quad .  
\end{equation}
Here  $\bar{x}$ is the sample mean 
defined as 
\begin{equation}
\bar{x} =\frac{1}{n} \sum_{i=1}^{n} x_i
\quad ,
\label{xmsample}
\end{equation}
formula which is useful  to derive the variance of the sample
\begin{equation}
Var =\frac{1}{n-1} \sum_{i=1}^{n} (x_i-{\bar x})^2
\quad ,
\label{varsample}
\end{equation}
where   $x_i$  are the n-data, see\cite{press}.
An example of random generation of points 
is reported in Figure \ref{df_maxwellrel}
where we imposed $T=1$ and we found $T=1.0397 $ 
from the generated random sample.

\subsection{Variable velocity} 

\begin{figure*}
\begin{center}
\includegraphics[width=5cm]{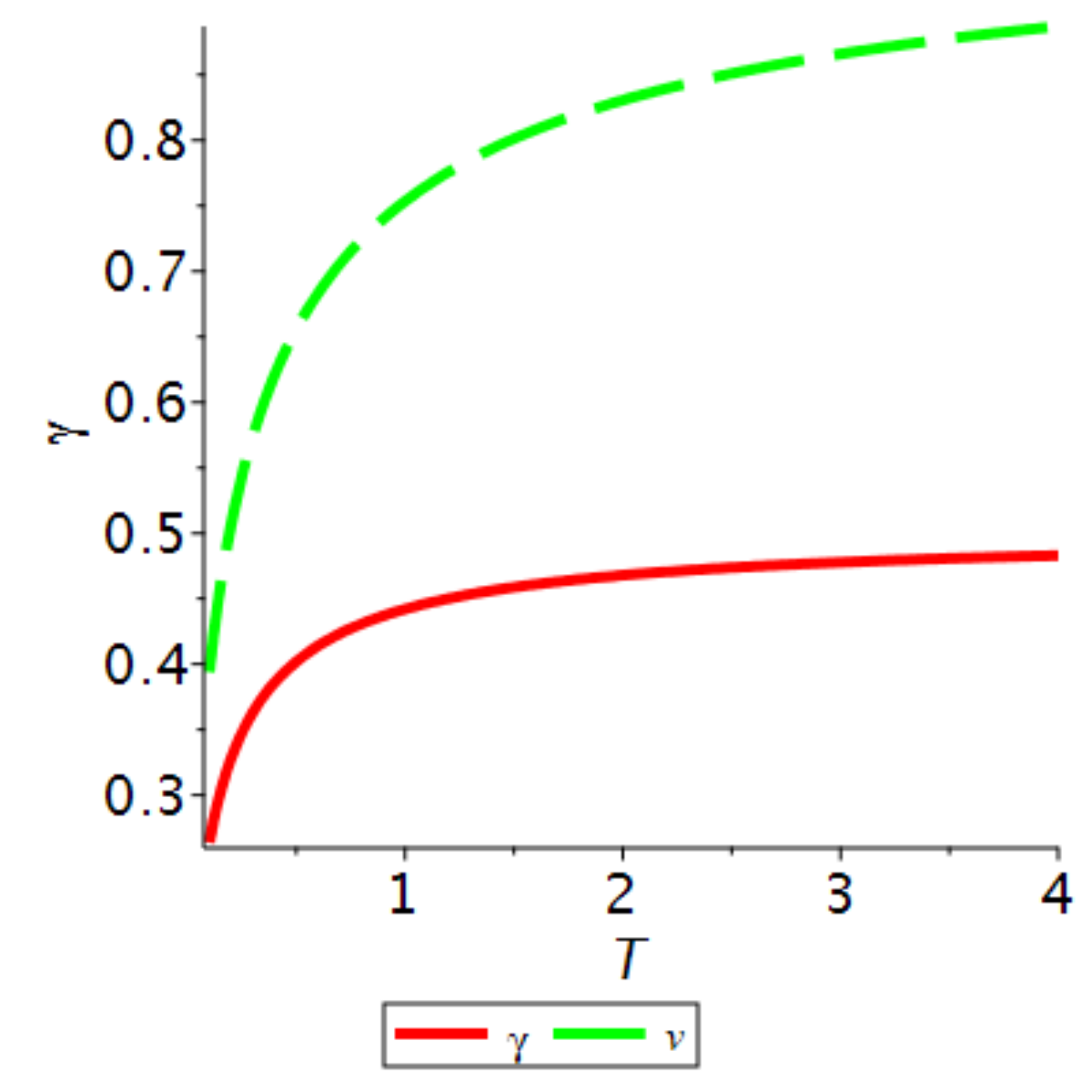}
\end {center}
\caption
{
The 
position  of the mode in the two PDFs:
gamma variable (red full line) and v variable (dashed green line).  
}
\label{modetwo}
    \end{figure*}

We now return  to the variable velocity, 
the PDF is 
\begin{equation}
f_r(v;T)=
\frac
{
32\,\sqrt {-{\frac {{v}^{2}}{{v}^{2}-1}}}{{\rm e}^{{\frac {\sqrt {-{v}
^{2}+1}-1}{\sqrt {-{v}^{2}+1}T}}}}\sqrt {-{v}^{2}+1}{T}^{3}v
}
{
{{\rm e}^{{T}^{-1}}}
G^{3, 0}_{1, 3}\left(1/4\,{T}^{-2}\, \Big\vert\,^{1}_{-1/2, -1, -3/2}\right)
}
\quad ,
\end{equation}
where $v$ is expressed in $c=1$ units.
The mode is a  solution  of a  sextic equation, see \cite{Hagedorn2000}, 
in $v$ 
\begin{equation}
-4\,{T}^{2}{v}^{6}+12\,{T}^{2}{v}^{4}-{v}^{4}-12\,{T}^{2}{v}^{2}+4\,{T
}^{2}
=0
\quad , 
\end{equation}
which has the following real solution       
\begin{eqnarray}
mode=
\frac{1}{6}\,\Bigg ( 3\,{\frac {\sqrt [3]{24\,\sqrt {3}\sqrt {27\,{T}^{2}-1}{T}
^{3}-216\,{T}^{4}+36\,{T}^{2}-1}}{{T}^{2}}}
\nonumber \\
-3\,{\frac {24\,{T}^{2}-1}{
{T}^{2}\sqrt [3]{24\,\sqrt {3}\sqrt {27\,{T}^{2}-1}{T}^{3}-216\,{T}^{4
}+36\,{T}^{2}-1}}}+3\,{\frac {12\,{T}^{2}-1}{{T}^{2}}}\Bigg )^{1/2}
\quad  .
\end{eqnarray}
The position of the mode for the PDF in $v$ is different 
from  that one in $\gamma$, see Figure \ref{modetwo}.

At the moment of writing the other statistical parameters cannot 
be presented in a closed form.

\section{The Maxwell J{\"u}ttner distribution}

\label{secmj}
\begin{figure*}
\begin{center}
\includegraphics[width=5cm]{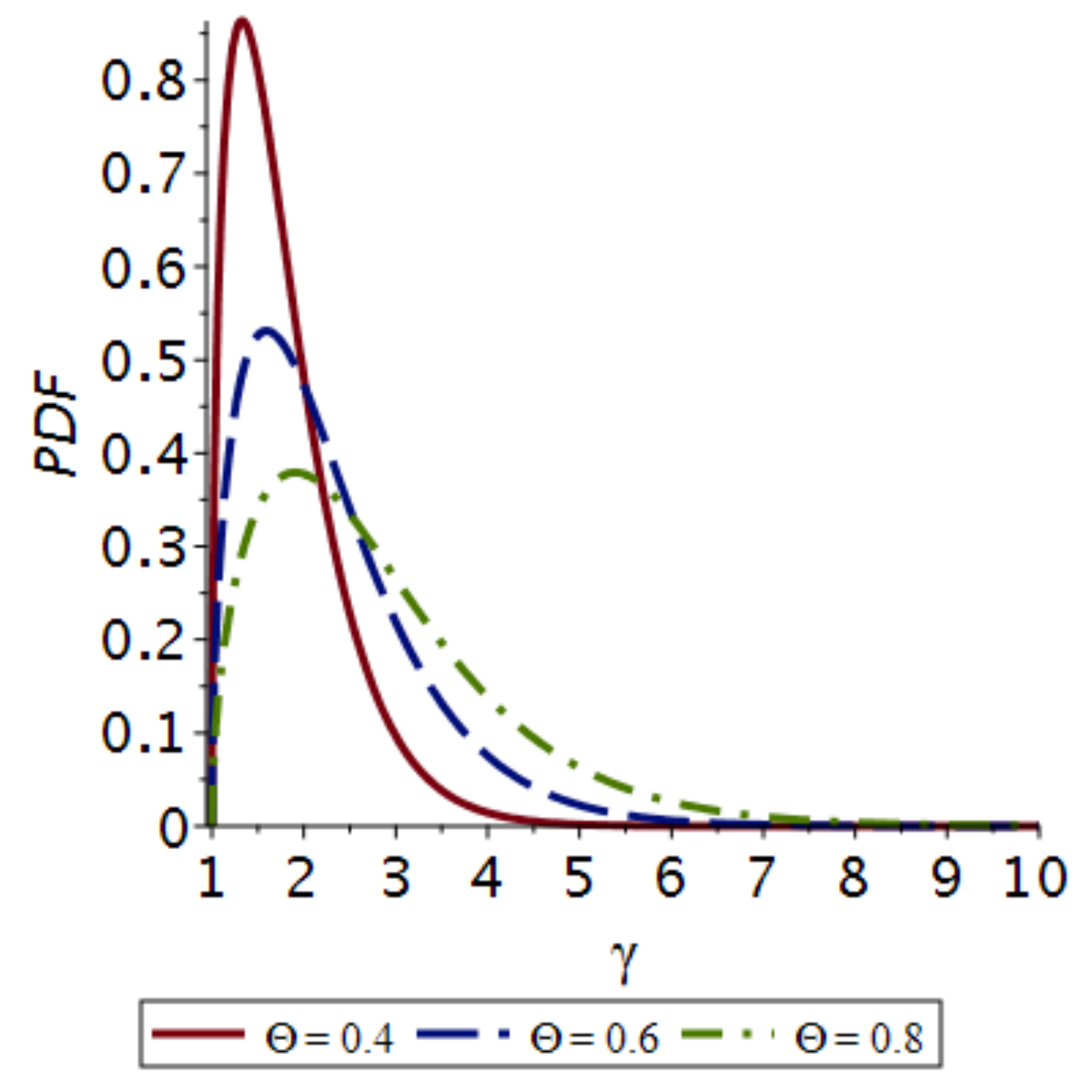}
\end {center}
\caption
{
The PDF of the MJ  as function of
$\gamma$ for different values of $\Theta$.
}
\label{maxjut3}
    \end{figure*}

\begin{figure*}
\begin{center}
\includegraphics[width=5cm]{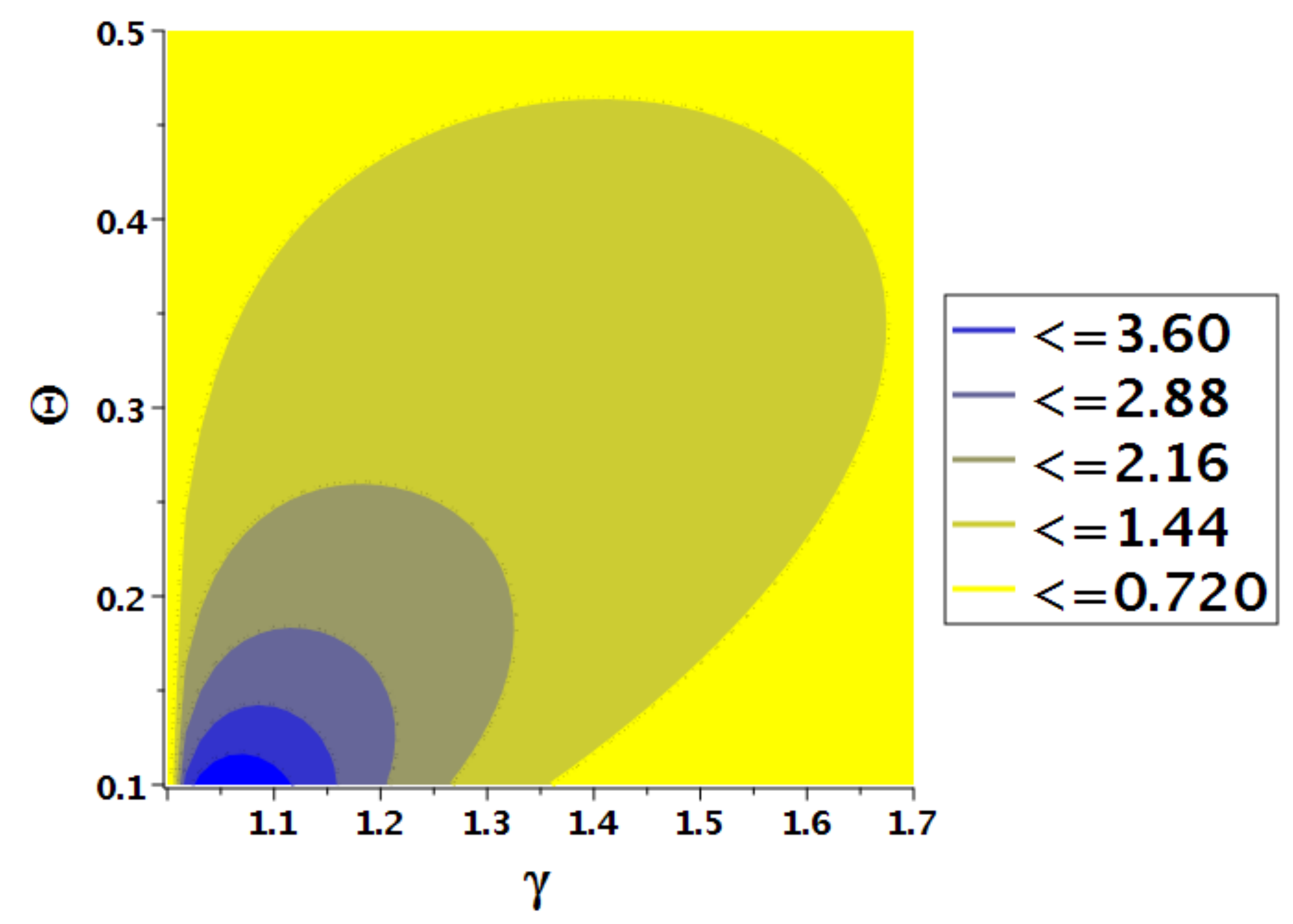}
\end {center}
\caption
{
A 2-D contour for
the MJ PDF   as function of
$\gamma$ and  $\Theta$.
}
\label{mjpdismap}
    \end{figure*}

The PDF for the  Maxwell J{\"u}ttner (MJ) distribution 
is 
\begin{equation}
f_{MJ} (\gamma;\Theta)
=
\frac
{
\gamma\,\sqrt {{\gamma}^{2}-1}{{\rm e}^{-{\frac {\gamma}{\Theta}}}}
}
{
\Theta\,{{\sl K}_{2}\left(\frac{1}{\Theta}\right)}
}
\quad ,
\label{pdfmaxwelljutner}
\end{equation}
where $\Theta=\sqrt{\frac{kT_{MB}}{m\,c^2}}$, 
$m$ is  the mass of the gas molecules,
$k$ is the Boltzmann constant, 
$T_{MB}$ is the usual thermodynamic temperature
and 
${\sl K}_2(x)$ is the Bessel function of second kind,
see \cite{Juttner1911,Synge1957,Livadiotis2016,Tsouros2017}.
Figure \ref{maxjut3}  reports the above PDF
for three different values of $\Theta$
and 
Figure \ref{mjpdismap}  displays  the  PDF
as a 2-D contour.

The average  value is
\begin{equation}
\mu(\Theta) =
\frac
{
-2\,{\Theta}^{2}
G^{2, 1}_{1, 3}\left(\frac{1}{4\Theta^2}\, \Big\vert\,^{1}_{3/2, -1/2,
- 2}\right)
}
{
{{\sl K}_{2}\left(\frac{1}{\Theta}\right)}
}
\end{equation}
and the variance is
\begin{eqnarray}
\sigma^2(\Theta)=
\frac
{
1
}
{
{\Theta}^{2} \left( {{\sl K}_{2}\left(\frac{1}{\Theta}\right)} \right) ^{
2}
}
\Bigg (
-4\,{\Theta}^{5} \Big( 2\,{{\sl K}_{1}\left({\Theta}^{-1}\right)}
G^{2, 1}_{1, 3}\left(1/4\,{\Theta}^{-2}\, \Big\vert\,^{1}_{5/2, -1/2,
- 2}\right)
\Theta
\nonumber 
\\
+ \left( 
G^{2, 1}_{1, 3}\left(1/4\,{\Theta}^{-2}\, \Big\vert\,^{1}_{3/2, -1/2,
2}\right)
 \right) ^{2}\Theta+{{\sl K}_{0}\left({\Theta}^{-1}\right)}
G^{2, 1}_{1, 3}\left(1/4\,{\Theta}^{-2}\, \Big\vert\,^{1}_{5/2, -1/2,
2}\right)
 \Big) 
\Bigg )
\quad .
\end{eqnarray}
The mode can be found by solving  the following 
cubic equation 
\begin{equation}
\frac {d}{d\,\gamma}
f_{MJ} (\gamma;\Theta)
\propto 
-{\gamma}^{3}+2\,\Theta\,{\gamma}^{2}+\gamma-\Theta=0
\quad .
\end{equation}
The real solution is         
\begin{eqnarray}
mode =
\frac
{
1
}
{
6\,\sqrt [3]{-36\,\Theta+64\,{\Theta}^{3}+12\,\sqrt {-96\,{\Theta}^{4}
-39\,{\Theta}^{2}-12}}
}
\times
\nonumber \\
\Bigg (
 \left( -36\,\Theta+64\,{\Theta}^{3}+12\,\sqrt {-96\,{\Theta}^{4}-39\,
{\Theta}^{2}-12} \right) ^{{\frac{2}{3}}}
\nonumber 
\\
+4\,\Theta\,\sqrt [3]{-36\,
\Theta+64\,{\Theta}^{3}+12\,\sqrt {-96\,{\Theta}^{4}-39\,{\Theta}^{2}-
12}}+16\,{\Theta}^{2}+12
\Bigg )
\quad .
\end{eqnarray}
The asymptotic expansion of order 10  for the PDF is
\begin{equation}
f_{MJ} (\gamma;\Theta)
\sim
\frac 
{1}
{
\Theta\,{{\sl K}_{2}
\left(
\frac
{1}
{\Theta}
\right)}
} 
\frac
{
 \left( 128\,{\gamma}^{8}-64\,{\gamma}^{6}-16\,{\gamma}^{4}-8\,{\gamma
}^{2}-5 \right) {{\rm e}^{-{\frac {\gamma}{\Theta}}}}
}
{
128\,{\gamma}^{6}
}
\quad .
\end{equation}
The DF is evaluated with the following integral
\begin{equation}
F_{MJ} (\gamma;\Theta)=\int_1^{\gamma}  f_{MJ} (\gamma;\Theta)\, d\gamma
\quad ,
\end{equation}
which cannot be expressed in terms of special functions.

\begin{figure*}
\begin{center}
\includegraphics[width=5cm]{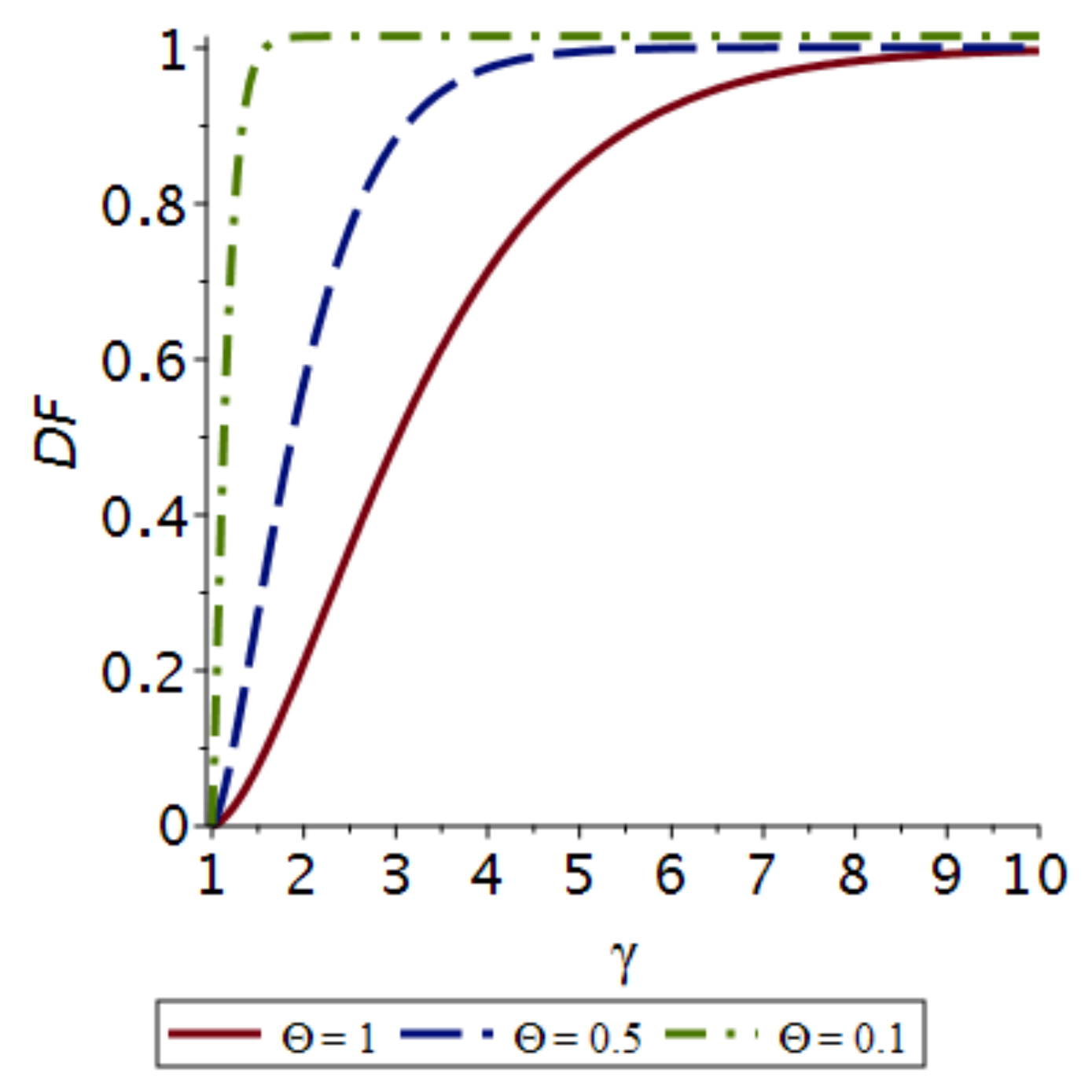}
\end {center}
\caption
{
The numerical MJ DF   as function of
$\gamma$ for different values of $\Theta$.
}
\label{dfmbasympt}
    \end{figure*}

\begin{figure*}
\begin{center}
\includegraphics[width=5cm,angle=-90]{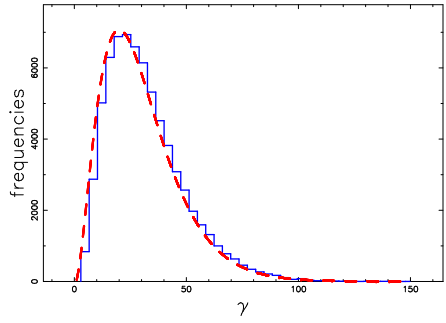}
\end {center}
\caption
{
The PDF  for 70000 random points
generated according to the MJ (blue steps) 
and the theoretical PDF       (red dashed line),   
see formula (\ref{pdfmaxwelljutner}).
}
\label{pdf_maxwellreljut}
    \end{figure*}

We now present some approximations for the distribution function
A {\it first} approximation is given by a series expansion 
when, ad example , $\Theta=1$
\begin{equation}
F_{MJ} (\gamma;1)= 
\frac{1}
{
{\sl K}_{2}(1)
}
\Bigg (
{{\sl K}_{2}\left(1\right)} +\sqrt {\pi}\sum _{m=0}^{\infty }{\frac { \left(
-1 \right) ^{1+m}
\Gamma \left( 3-2\,m,\gamma \right) }{\Gamma \left( 1+m \right) 
  \Gamma \left( {\frac{3}{2}}-m \right)  }}
\Bigg ) \quad ,
\end{equation}
which has a percent error less $< 0.6 \% $ in interval
$1.1 < \gamma <10$ when  $T=1$.
A {\it second} approximation is given by an asymptotic 
expansion of order 50 for the PDF followed by
the integration, see Figure \ref{dfmbasympt}.
The  parameter $\Theta$ can be derived from the  experimental
sample once the average value is 
modeled by a  Pade approximant $[2,2]$ and the inverse function 
is derived
\begin{equation}
\Theta=
0.1661\,{\it \bar{x}}- 0.3085+ 1.36051 \times 10^{-10}\,\sqrt {
 1.4908 \times10^{18}\,{{\it \bar{x}}}^{2}+ 5.913\times10^{18}\,{\it \bar{x}
}- 6.5835\times10^{18}}
\label{theta_da_ave}
\quad .
\end{equation}
An analogous formula allows to derive  $\Theta$ from the variance $Var$
of the sample
\begin{eqnarray}
\Theta=
\frac{1}{4} \times
\nonumber \\
{\frac { 1.818\times 10^{10}  {\it Var}+ 5.972\times 10^{11}+
 5 \sqrt {{ 2.277\times 10^{20}}  {{\it Var}}^{2}+{
 7.814\times 10^{23}}  {\it Var}-{ 3.597\times 10^{22
}}}}{ 5.436\times 10^{8}  {\it Var}+ 1.978\times10^{12}}}
\quad .
\label{theta_da_var}
\end{eqnarray}
An example of random generation of points 
is reported in Figure \ref{pdf_maxwellreljut}
where we imposed $T=10$ and we found 
$T=9.97 $ from formula (\ref{theta_da_ave}) 
and
$T=9.98$ from formula  (\ref{theta_da_var}).

\subsection{Variable $\beta$ } 

\begin{figure*}
\begin{center}
\includegraphics[width=5cm]{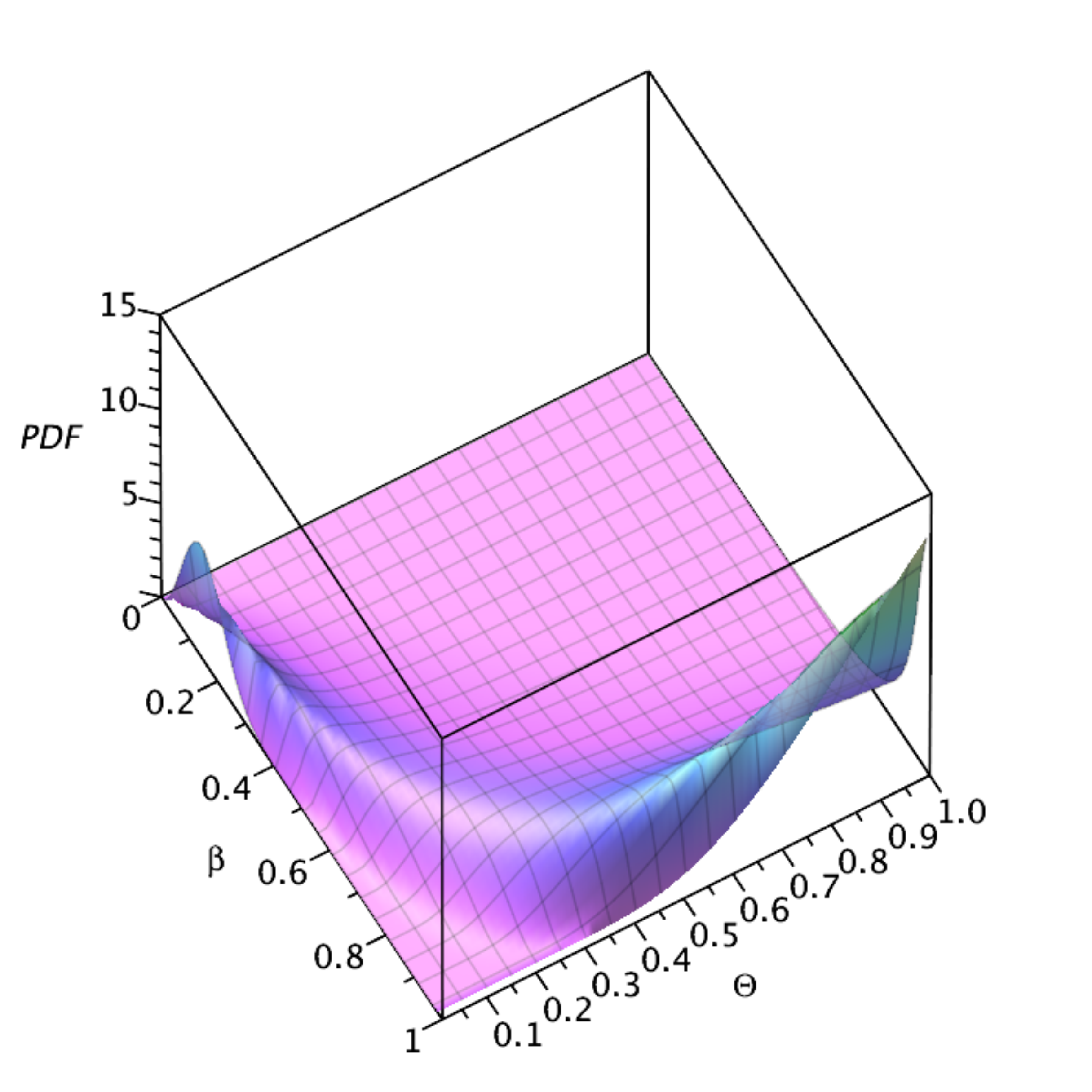}
\end {center}
\caption
{
The 3D surface of the MJ PDF as function of $\Theta$ and $\beta$.
}
\label{mjbetadis}
    \end{figure*}

\begin{figure*}
\begin{center}
\includegraphics[width=5cm]{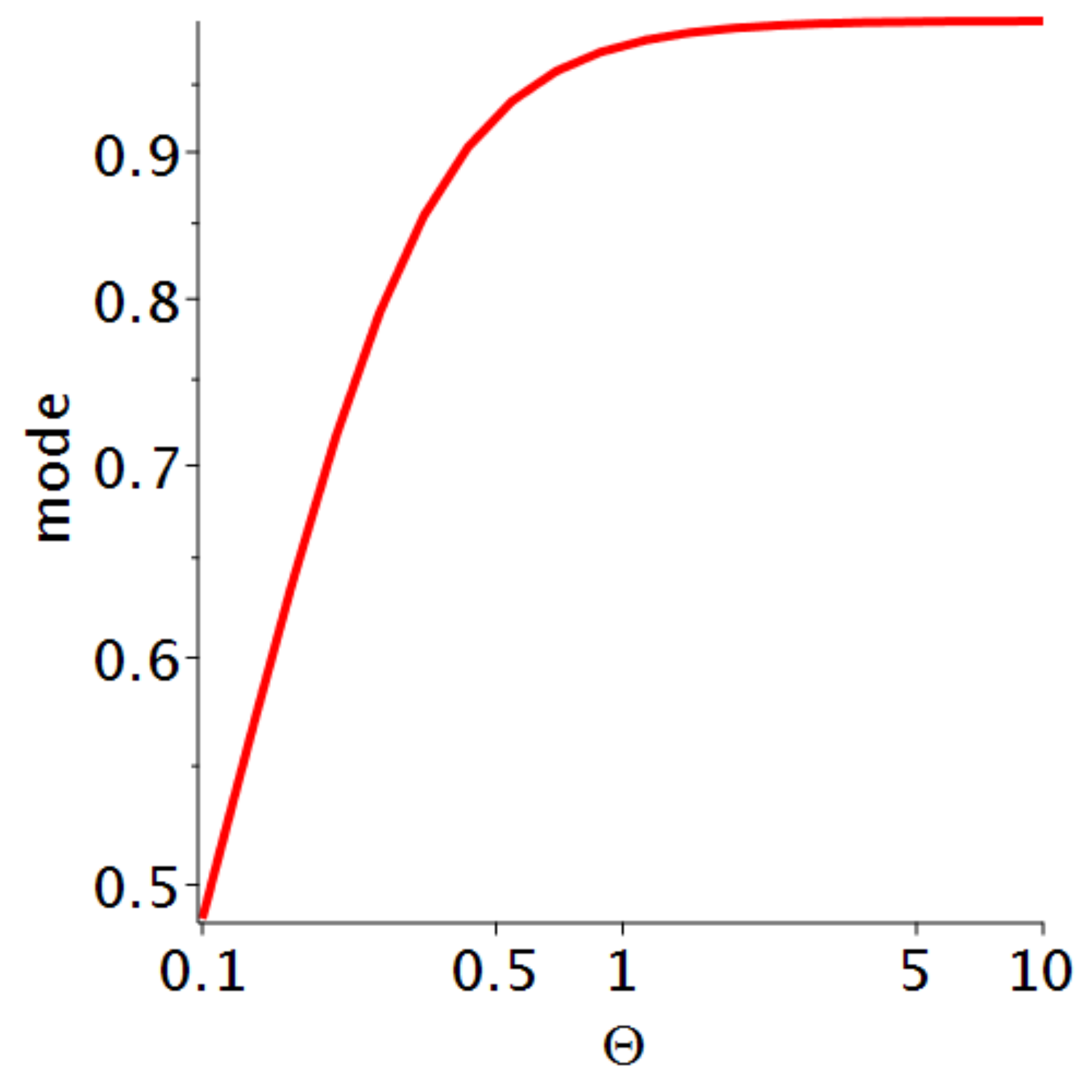}
\end {center}
\caption
{
The mode of the MJ distribution  for different values of $\Theta$.
}
\label{mjmodetheta}
    \end{figure*}

We now change the variable of integration $\gamma$ in $\beta=\frac{v}{c}$,
the  PDF of the MJ is
\begin{equation}
f_{MJ} (\beta;\Theta)
=
\frac
{
\sqrt { \left( 1-{\beta}^{2} \right) ^{-1}-1}{{\rm e}^{-{\frac {1}{
\Theta}{\frac {1}{\sqrt {1-{\beta}^{2}}}}}}}\beta
}
{
\left( {1-\beta}^{2} \right) ^{2}\Theta\,{{\sl K}_{2}\left(\frac{1}{\Theta}
\right)}
}
\quad ,
\end{equation}
where $0\leq\beta\leq 1$, see Figure \ref{mjbetadis}.
We have only one  analytical result, the mode,
which is found solving the following equation in $\beta$
\begin{equation}
-3\, \left( \beta-1 \right) ^{3} \left( \beta+1 \right) ^{3} \left( 
\Theta\, \left( {\beta}^{2}+2/3 \right) \sqrt {-{\beta}^{2}+1}-1/3\,{
\beta}^{2} \right) {{\rm e}^{-{\frac {1}{\sqrt {-{\beta}^{2}+1}\Theta}
}}}{\beta}^{2}=0
\quad .
\end{equation}
As an example  when $\Theta=0.1$  the mode is at $\beta=0.4866$
and Figure \ref{mjmodetheta} reports the mode as function 
of $\Theta$.

\begin{figure*}
\begin{center}
\includegraphics[width=5cm]{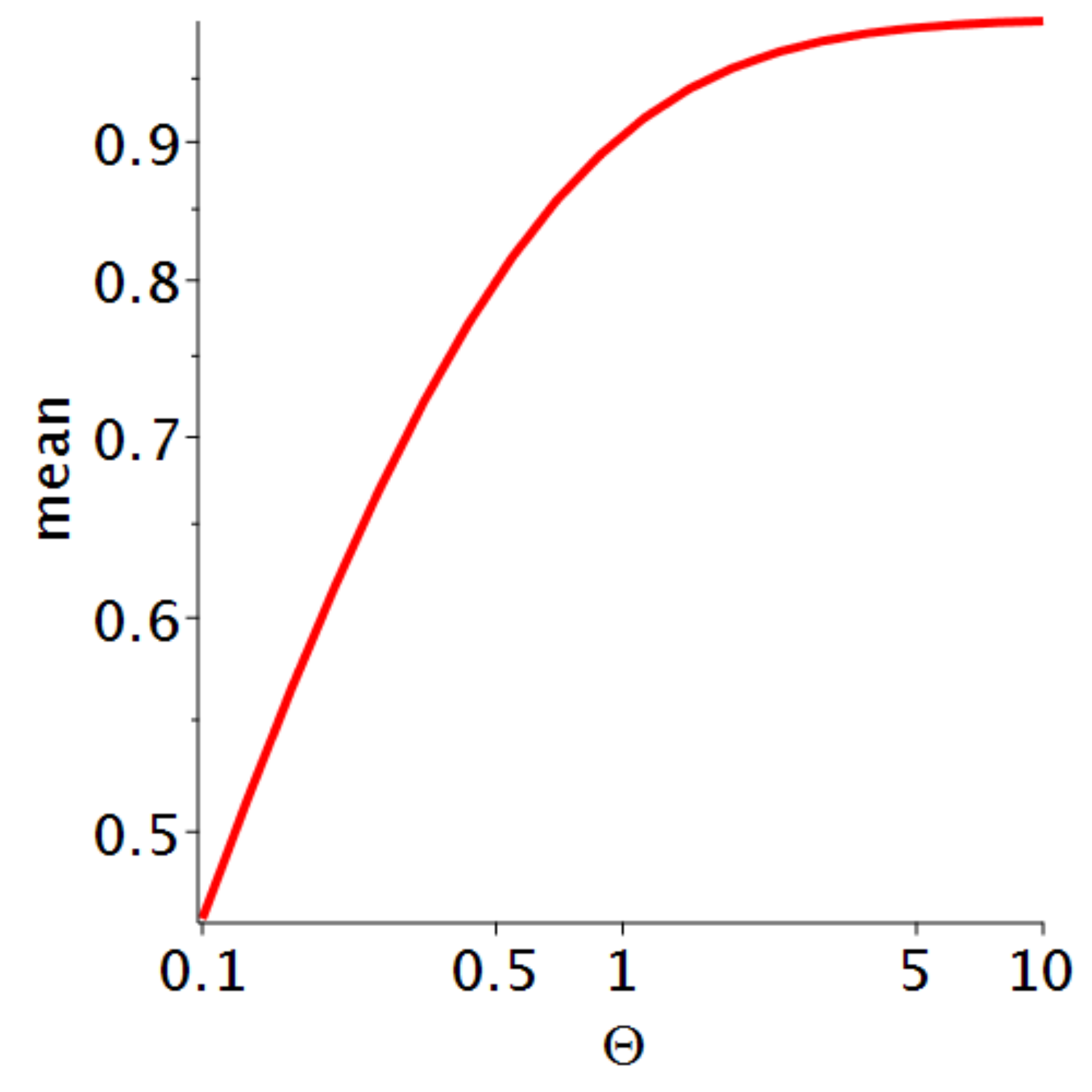}
\end {center}
\caption
{
The mean of the MJ distribution  for different values of $\Theta$.
}
\label{mjmean}
    \end{figure*}

\begin{figure*}
\begin{center}
\includegraphics[width=5cm]{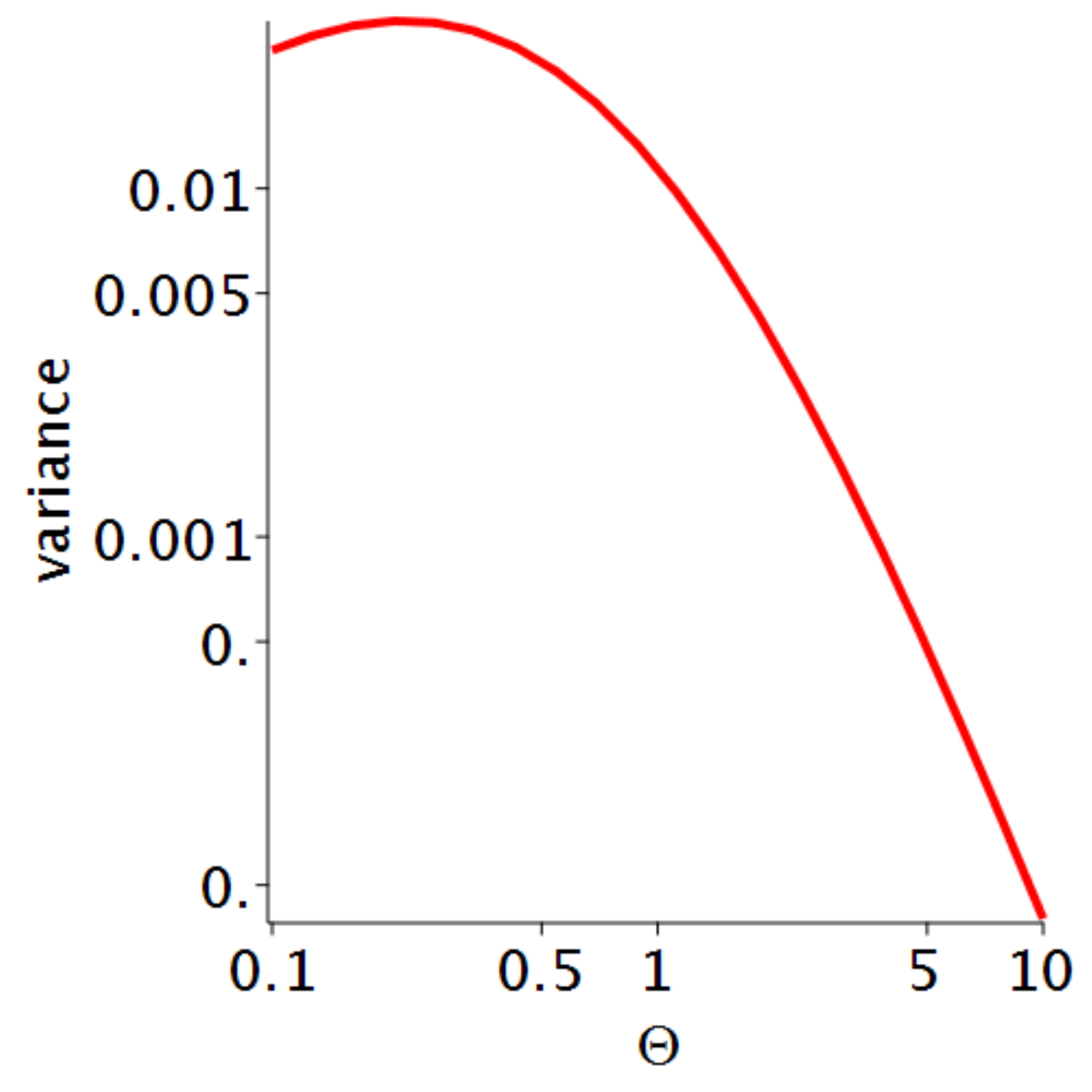}
\end {center}
\caption
{
The variance  of the MJ distribution  for different values of $\Theta$.
}
\label{mjvariance}
    \end{figure*}

The mean and the variance of the MJ distribution does not have
an analytical expression and they are reported 
in a numerical way, see Figures \ref{mjmean}  and \ref{mjvariance}.

\begin{figure*}
\begin{center}
\includegraphics[width=5cm]{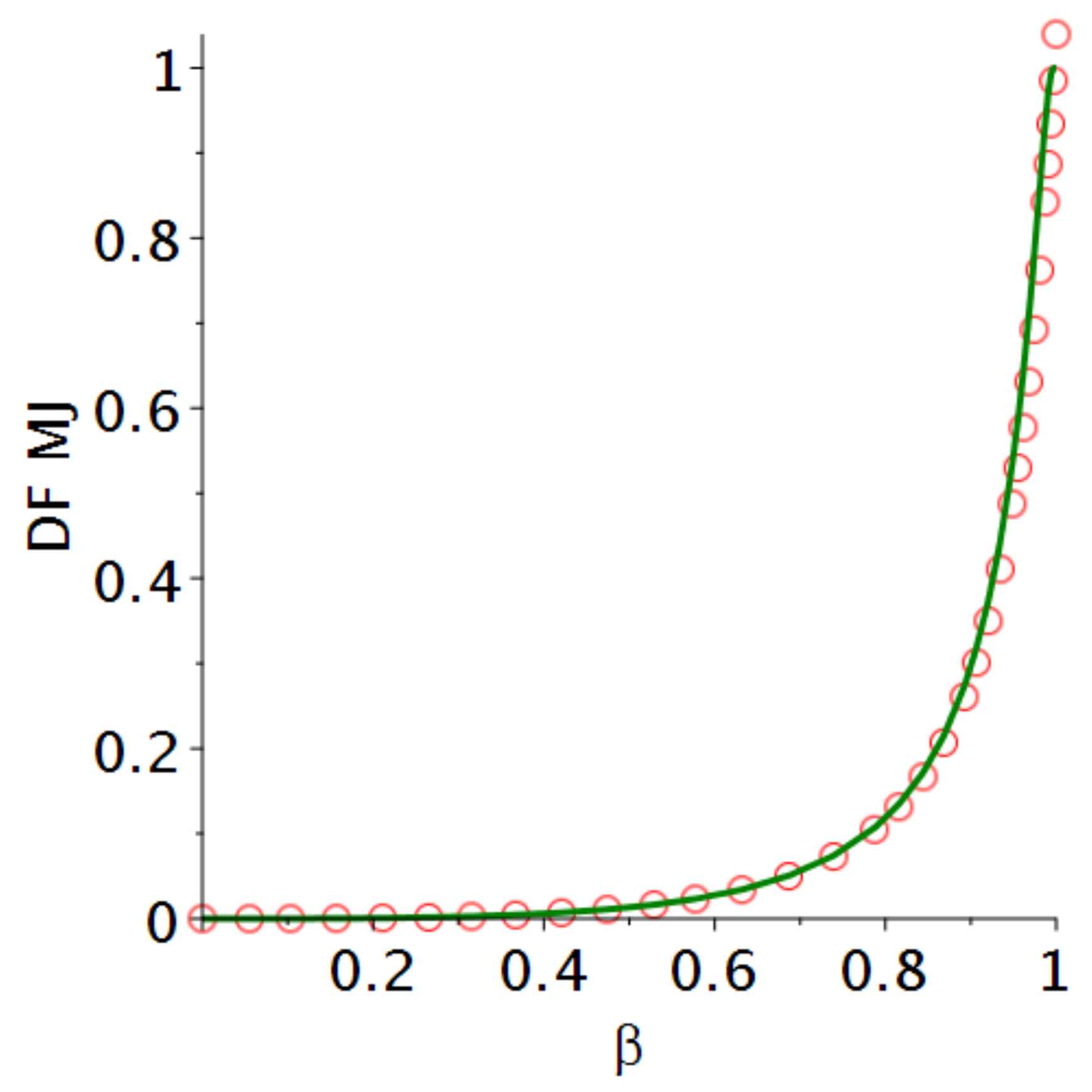}
\end {center}
\caption
{
The numerical DF  of the MJ distribution  (red circles)
and the Riemann approximation (green line) 
as function of $\beta$.
}
\label{mjdfriemann}
    \end{figure*}

The DF of the MJ is given  by the following integral
\begin{equation}
F_{MJ} (\beta;\Theta) = \int_0^{\beta}  f_{MJ} (\beta;\Theta)\, d\beta
\quad ,
\end{equation}
with $\beta$ in [0,1] 
which does not have an analytical expression.
An  approximation is given by the Riemann sums, see \cite{Anton2012}, 
when $\Theta=1$ 
\begin{eqnarray}
F_{MJ} (\beta;\Theta) = \nonumber\\  
\frac
{
\beta\sum _{{\it i}=0}^{9}{\frac {\beta}{10\,{{\sl K}_{2}\left(1\right)}}
\sqrt { \left( -{\frac {{\beta}^{2}}{100} \left( {\it i}+{\frac{1}{2}}
 \right) ^{2}}+1 \right) ^{-1}-1}{{\rm e}^{-{\frac {1}{\sqrt {-{\frac 
{{\beta}^{2}}{100} \left( {\it i}+{\frac{1}{2}} \right) ^{2}}+1}}}}}
 \left( {\it i}+{\frac{1}{2}} \right)  \left( -{\frac {{\beta}^{2}}{100}
 \left( {\it i}+{\frac{1}{2}} \right) ^{2}}+1 \right) ^{-2}}
}
{
10
}
\quad ,
\end{eqnarray}
see Figure \ref{mjdfriemann}.
The above DF  has a maximum percentage error of $\approx 10\%$ at 
$\beta=1$.

\section{The astrophysical applications}

\label{secastro}
This section  reviews the synchrotron emissivity for a single
relativistic electron,
derives the spectral synchrotron emissivity for
the two relativistic distributions here analyzed 
and models the observed synchrotron emission 
in some astrophysical sources.

\subsection{Synchrotron emissivity}

\begin{figure*}
\begin{center}
\includegraphics[width=5cm]{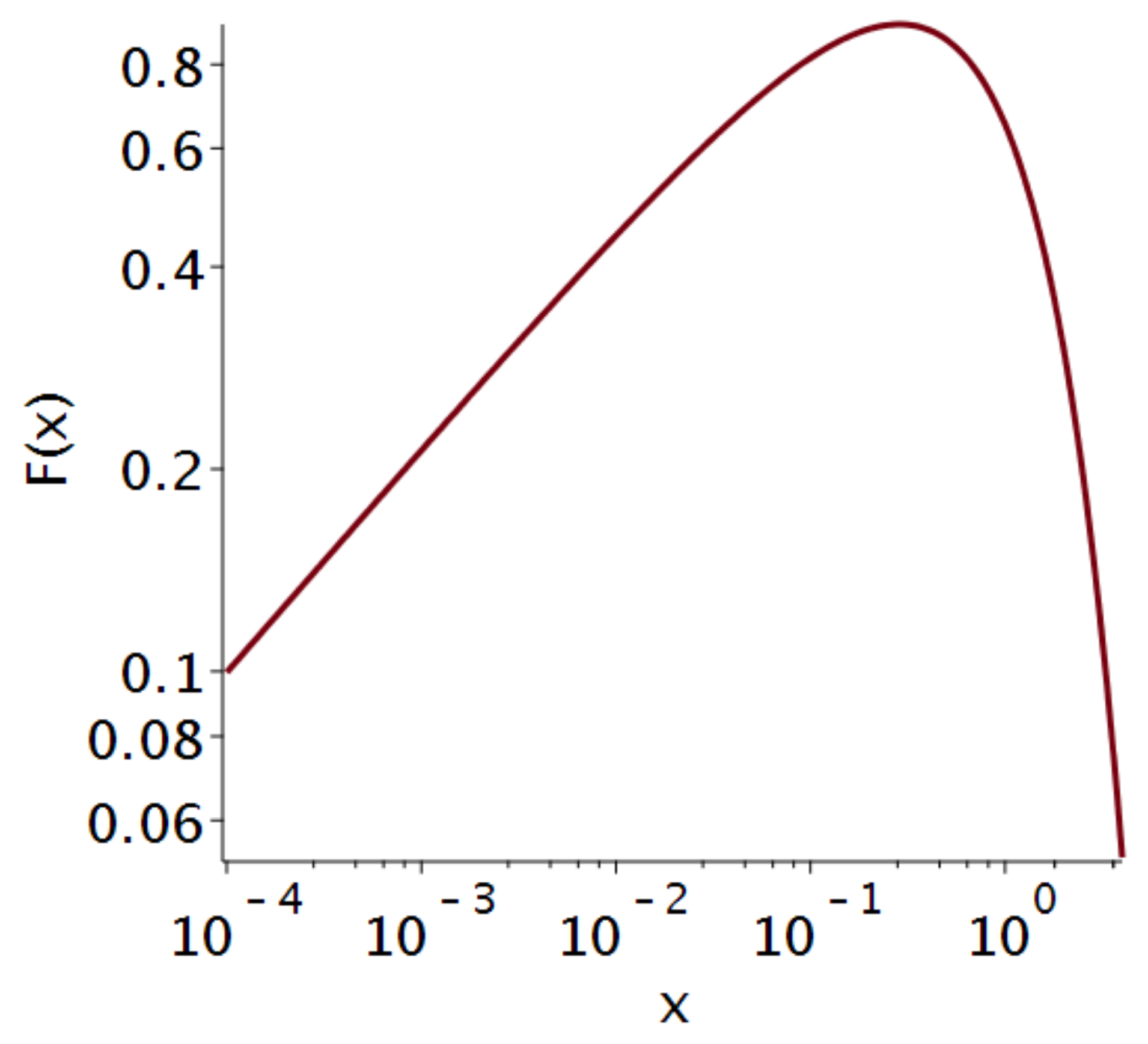}
\end {center}
\caption
{
F(x) as function of $x$ with logarithmic axes.
}
\label{loglogplotfx}
    \end{figure*}

The synchrotron emissivity of a single electron 
is 
\begin{equation}
\frac
{
\sqrt{3} e^3 B \sin(\alpha) 
}
{
8 \pi^2 \epsilon_0 c m_e
}
F(x)
\quad ,
\end{equation}
where,  according to eqn.(8.58) in  
\cite{Longair2011},
$e$  is the electron charge,
$B$  is the magnetic field,
$\alpha$ is the pitch angle,
$\epsilon_0$ is the permittivity of free space,
$c$ is the light velocity,
$m_e$ is the electron mass,
$x=$ is the ratio of the  angular frequency ($\omega$)
to the  critical angular frequency ($\omega_c$) 
and  
\begin{equation}
F(x) = x \int_x^{\infty} K_{5/3} (z) dz
\end{equation}
where  $K_{5/3} (z)$ is the modified Bessel function of second kind 
with order 5/3 \cite{Abramowitz1965,NIST2010}.
The modified Bessel function is also known 
as Basset function, modified Bessel function of the third kind 
 or Macdonald function see pag. 527 in \cite{Oldham2010}. 
The above function has  the following analytical
expression
\begin{equation}
F(x)
=
-{\frac {9\,\sqrt {3}\sqrt [3]{2}\pi}{320\,\Gamma \left( 2/3 \right) }
{x}^{{\frac{11}{3}}}
{\mbox{$_1$F$_2$}({\frac{4}{3}};\,{\frac{7}{3}},{\frac{8}{3}};\,{\frac
{{x}^{2}}{4}})}
}-{\frac {x\sqrt {3}\pi}{3}}+\sqrt [3]{x}{2}^{{\frac{2}{3}}}\Gamma
 \left( {\frac{2}{3}} \right) 
{\mbox{$_1$F$_2$}(-{\frac{1}{3}};\,-{\frac{2}{3}},{\frac{2}{3}};\,{\frac
{{x}^{2}}{4}})}
\quad ,
\end{equation} 
where ${\2F1(a,b;\,c;\,v)}$ is a regularized hypergeometric function 
\cite{Abramowitz1965,Seggern1992,Thompson1997,NIST2010}.  
Figure \ref{loglogplotfx} displays   $F(x)$ as function 
of $x$.

\subsection{The synchrotron  relativistic MB distribution} 

We start from the PDF for the relativistic MB distribution
as represented by equation (\ref{pdfgammarel}) 
and we perform the following {\it first} 
change of variable
\begin{equation}
\gamma = \frac{E}{m_ec^2}
\quad ,
\end{equation}
where $E$ is the relativistic energy.
The resulting PDF in relativistic energy
is 
\begin{equation}
f_r(E;T)
=
\frac
{
32\,\sqrt {{\frac {{E}^{2}}{{m_{{e}}}^{2}{c}^{4}}}-1}{{\rm e}^{{\frac 
{1}{T} \left( 1-{\frac {E}{m_{{e}}{c}^{2}}} \right) }}}{T}^{3}{m_{{e}}
}^{3}{c}^{6}
}
{
{E}^{4}{{\rm e}^{{T}^{-1}}}
G^{3, 0}_{1, 3}\left({\frac {1}{4\,{T}^{2}}}\,
\Big\vert\,^{1}_{-{\frac{1}{2}}, -1, -{\frac{3}{2}}}\right)
}
\quad .
\end{equation}
A {\it second} 
change of variable
is 
\begin{equation}
E = \sqrt {{\frac {\nu}{\nu_{{g}}}}}m_{{e}}{c}^{2}
\quad ,
\end{equation}
produces
\begin{equation}
f_r(\nu;T,\nu_{{g}})=
\frac
{
16\,\sqrt {{\frac {\nu}{\nu_{{g}}}}-1}{{\rm e}^{{\frac {1}{T} \left( 1
-\sqrt {{\frac {\nu}{\nu_{{g}}}}} \right) }}}{T}^{3}\nu_{{g}}
}
{
{\nu}^{2}{{\rm e}^{{T}^{-1}}}
G^{3, 0}_{1, 3}\left({\frac {1}{4\,{T}^{2}}}\,
\Big\vert\,^{1}_{-{\frac{1}{2}}, -1, -{\frac{3}{2}}}\right)
\sqrt {{\frac {\nu}{\nu_{{g}}}}}
}
\quad ,
\end{equation}
where 
\begin{equation}
\nu_g = \frac{e B}{2\pi m_e}
\quad .
\end{equation}
We know   that $\nu_g=2.799249\,10^{12} \,B $ 
where $B$ is the magnetic field expressed in gauss
and therefore the above PDF in frequency becomes 
\begin{equation}
f_r(\nu;T,B)
=
\frac
{
{ 7.49345\times 10^{19}}\,\sqrt {{ 3.57238\times 10^{-13}}\,{
\frac {\nu}{B}}-1}{{\rm e}^{{\frac {1}{T} 
\left( 1- 5.97694 \times 10^{-7}
\,\sqrt {{\frac {\nu}{B}}} \right) }}}{T}^{3}B
}
{
{\nu}^{2}{{\rm e}^{{T}^{-1}}}
G^{3, 0}_{1, 3}\left({\frac {1}{4\,{T}^{2}}}\,
\Big\vert\,^{1}_{-{\frac{1}{2}}, -1, -{\frac{3}{2}}}\right)
\sqrt {{\frac {\nu}{B}}}
}
\quad .
\end{equation}

\subsection{The synchrotron  Maxwell J{\"u}ttner distribution} 

We start from the PDF for the  Maxwell J{\"u}ttner distribution
as  given by equation (\ref{pdfmaxwelljutner}) 
and we perform  two changes in variable as in the
previous section.
The resulting PDF in relativistic energy
is 
\begin{equation}
f_{MJ} (E;\Theta)
=
\frac
{
E\sqrt {{\frac {{E}^{2}}{{m_{{e}}}^{2}{c}^{4}}}-1}{{\rm e}^{-{\frac {E
}{m_{{e}}{c}^{2}\Theta}}}}
}
{
{m_{{e}}}^{2}{c}^{4}\Theta\,{{\sl K}_{2}\left( \frac{1}{\Theta} \right)}
}
\quad .
\end{equation}
The second PDF  in  $\nu$  is  
\begin{equation}
f_{MJ} (\nu;\Theta,\nu_{{g}})
=
\frac
{
\sqrt {{\frac {\nu}{\nu_{{g}}}}-1}{{\rm e}^{-{\frac {1}{\Theta}\sqrt {
{\frac {\nu}{\nu_{{g}}}}}}}}
}
{
2\,\Theta\,{{\sl K}_{2}\left(\frac{1}{\Theta} \right)}\nu_{{g}}
}
\quad .
\end{equation}
The astrophysical PDF in frequency for  
the  Maxwell J{\"u}ttner distribution
is
\begin{equation}
f_{MJ} (\nu;\Theta,B)
=
\frac
{
{ 1.78619\times 10^{-13}}\,\sqrt {{ 3.57238\times 10^{-13}}\,{
\frac {\nu}{B}}-1}{{\rm e}^{- 5.97694 \times 10^{-7}   \,{\frac {1}{\Theta}
\sqrt {{\frac {\nu}{B}}}}}}
}
{
\Theta\,{{\sl K}_{2}\left( \frac{1}{\Theta}  \right)}B
}
\quad .
\label{pdfmjastro}
\end{equation}
The mismatch  between  measured flux in Jy 
and theoretical flux, $S_{theo}$,
can be obtained
introducing a multiplicative constant $C$
\begin{equation}
S_{theo} = C \times f_{MJ} (\nu;\Theta,B)
\quad .
\label{stheomj}
\end{equation}

\subsection{The spectrum of the radio-sources}

\begin{figure*}
\begin{center}
\includegraphics[width=5cm,angle=-90]{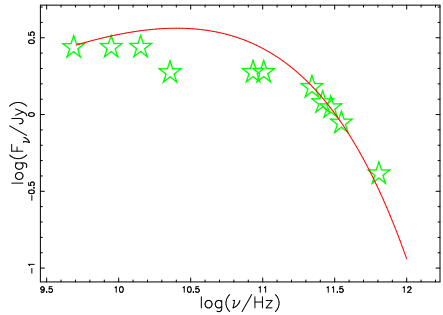}
\end {center}
\caption
{
Measured flux in Jy (green stars) of the quiet core of M87  
and the theoretical flux  (red line) for 
the  Maxwell J{\"u}ttner distribution 
as given by equation (\ref{stheomj}).
The numerical parameters are $B=10^{-5}\,gauss$,
$\Theta=30$ and $C=10^{12}\,Jy$.
}
\label{radiombrel}
    \end{figure*}

As a {\it first } example we analyze the spectrum of 
an extended region around M87, 
see as example Figure 1 in  \cite{Prieto2016} 
where the flux   in Jy as function of the
frequency is reported 
in the range  $9\times10^9 \,Hz <\nu< 2 \times 10^{18}\, Hz$.
Figure \ref{radiombrel} reports 
the measured and theoretical flux in the range 
$9\times10^9 \,Hz <\nu< 2 \times 10^{12}\, Hz$
for the quiet core of M87.

\begin{figure*}
\begin{center}
\includegraphics[width=5cm]{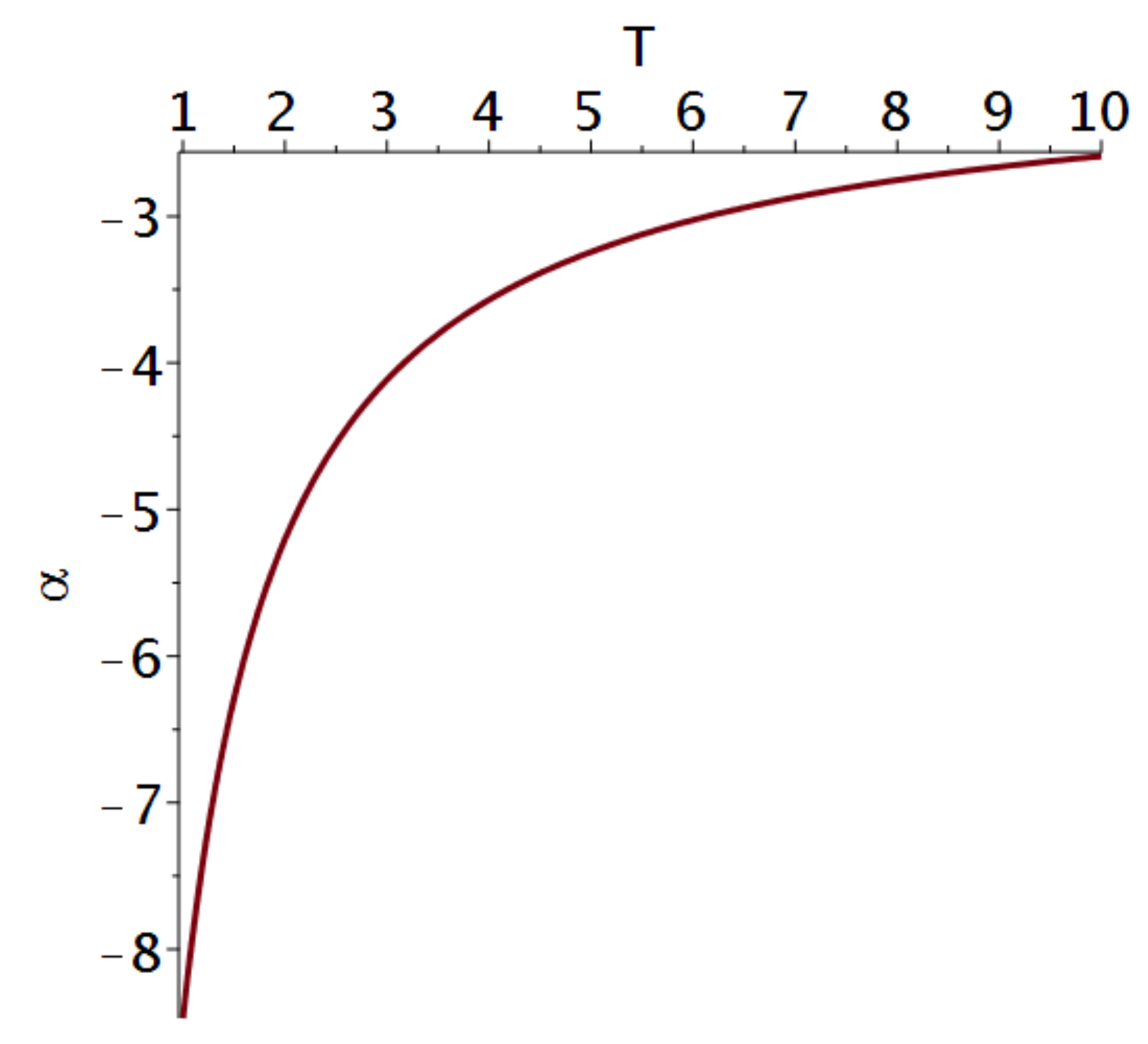}
\end {center}
\caption
{
The spectral index of the  relativistic MB as function of
$T$  when $B=1.0\times10^{-6}$.
}
\label{alphat}
    \end{figure*}

\begin{figure*}
\begin{center}
\includegraphics[width=5cm]{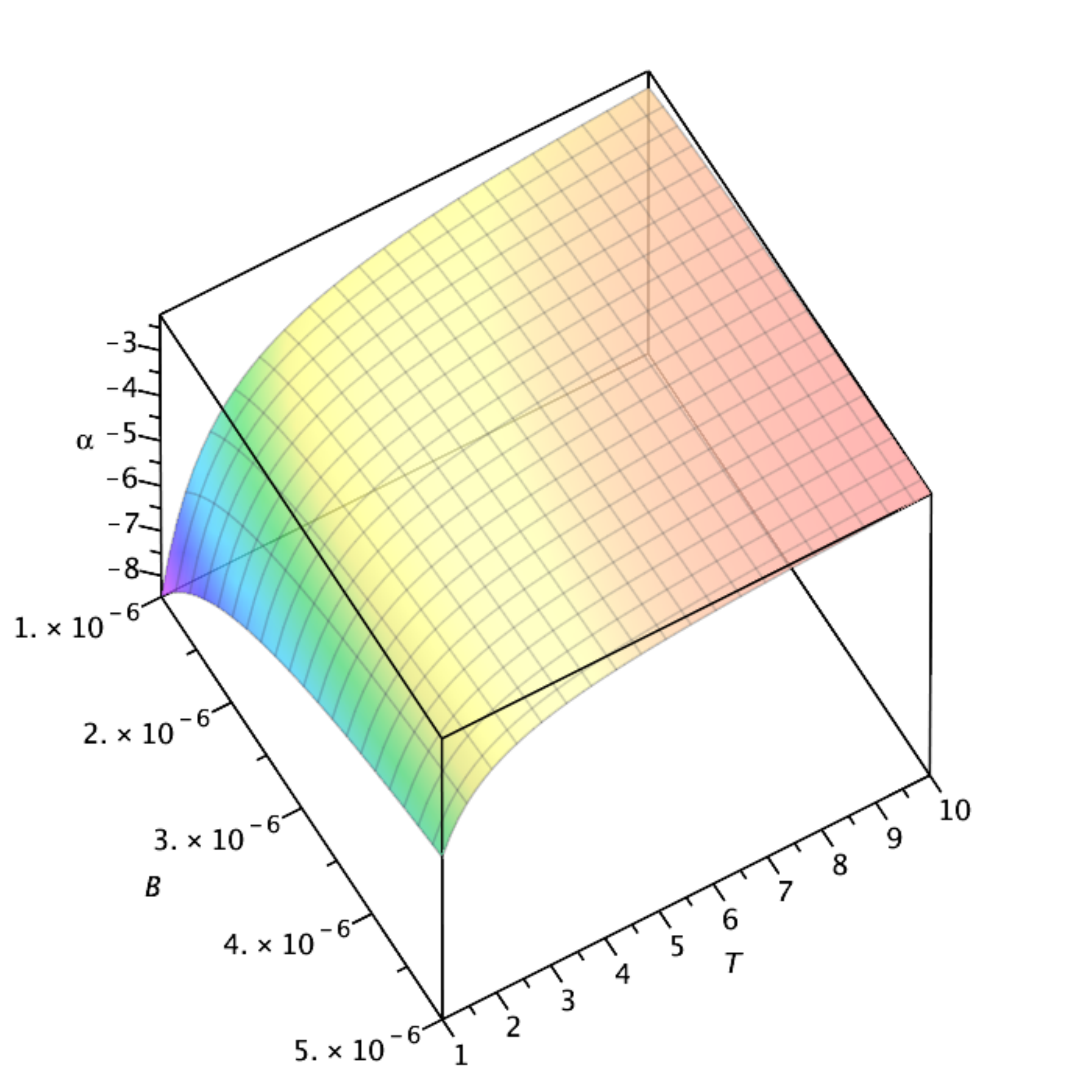}
\end {center}
\caption
{
The spectral index of the  relativistic MB as function of
$T$  and  $B$ in gauss.
}
\label{btalpha}
    \end{figure*}

A {\it second} example is given 
by the  radio sources with ultra steep spectra (USS) 
which are characterized by a spectral index, $\alpha$,
lower than -1.30 
when the radio flux, $S$, is proportional to $S^{\alpha}$,
see \cite{DeBreuck2000}.
As a  practical example  we select the
cluster Abell 1914 where  
the measured total flux densities
at 150 MHz and  1.4 GHz 
are $S_{150} = 4.68\,$Jy
and $S_{1.4}$  = 34.8\,mJy which means 
$\alpha =-2.17$.
We now evaluate the  theoretical spectral index 
of synchrotron emission for the relativistic MB distribution
between 150 MHz and 1.4 GHz when $B$ is fixed and $T$ variable, see
Figure \ref{alphat}
and Figure \ref{btalpha} when
$T$ and $B$ are both variables.
The two Figures above show that  the theoretical spectral index is 
always smaller than -2 which can be considered as an asymptotic
limit for high values of relativistic temperature.   
As an example when $B=1.0\times 10^{-5} gauss$ the 
spectral index is -2.17 when $T=10$.

\section{Conclusions}

The relativistic MB distribution has been derived  
in \cite{Claycomb2018} without any particular 
statistics: here we derived, when the main 
variable is the Lorentz factor $\gamma$,  
the constant of normalization,
the average value,
the second moment about the origin,
the variance,
the mode,
the asymptotic behavior,
an approximate expression for the average value as function of the
temperature and 
an inverted expression for the temperature as function of 
average value.

We derived the following statistical parameters of the  
MJ distribution when $\gamma$ is the main variable: average value,
variance,
mode,
asymptotic expansion,
two approximate expressions for the distribution function,
a first  evaluation of $\Theta$ from the average value
and  
a second evaluation of $\Theta$ from the variance. 

Following the usual argument which suggests a power law 
behavior for the spectral distribution of the 
synchrotron emission in presence 
of a power law distribution for the energy of the electrons   
we derived the spectral distribution for the
relativistic MB and  MJ distributions which
are now function of the selected generalized temperature 
and the magnetic field.
Two  astrophysical applications are given:
the spectral distribution of
emission in the core of M87 in the framework of the synchrotron 
emissivity and an  explanation for the 
steep spectra sources in the framework of the
synchrotron emissivity for the relativistic MJ distribution.
\providecommand{\newblock}{}

\end{document}